\documentstyle[aps,amssymb,preprint]{revtex}
\tightenlines  
  
\begin{document}

\title{Variable survival exponents in history-dependent random walks: 
hard movable reflector   
}  

\author{Ronald Dickman$^{1}$\footnote{dickman@fisica.ufmg.br}, 
Francisco Fontenele Araujo Jr.$^{1}$\footnote{ffaraujo@fisica.ufmg.br} 
and Daniel ben-Avraham$^{2}$\footnote{benavraham@clarkson.edu}}
\address{$^1$Departamento de F\'{\i}sica, ICEx,   
Universidade Federal de Minas Gerais,    
C.P. 702, 30.161-970,   
Belo Horizonte - Minas Gerais, Brazil\\
$^2$Physics Department, 
Clarkson University, Potsdam, NY 13699-5820}   

\date{\today}

\maketitle
  
\begin{abstract}  
We review recent studies demonstrating a nonuniversal (continuously
variable) survival exponent for history-dependent random walks,
and analyze a new example, the hard movable partial reflector.
These processes serve as a simplified models of infection in
a medium with a history-dependent susceptibility, and for 
spreading in systems with an infinite number of absorbing 
configurations. 
The memory may take the form of a history-dependent
step length, or be the result  
of 
a partial reflector whose position
marks the maximum distance the walker has ventured from the origin.
In each case, a process with memory is rendered
Markovian by a suitable expansion of the state space.
Asymptotic analysis of the probability 
generating function shows that, for large $t$,  
the survival probability decays as 
$S(t) \sim t^{-\delta}$, where $\delta$ varies
with the parameters of the model.
We report new results for a {\it hard} partial reflector, i.e., one that
moves forward only when the walker does. 
When the walker tries to jump to the site R occupied by the
reflector, it is reflected back with probability $r$,
and stays at R with probability $1-r$; only in the
latter case does the reflector move (R $\to $ R+1).
For this model, $\delta = 1/2(1\!-\!r)$, and 
becomes 
arbitrarily large as $r$ approaches 1.  This prediction is confirmed via
iteration of the transfer matrix, which also reveals slowly-decaying
corrections to scaling.
\end{abstract}  

  
\section{Introduction}  
\label{intro} 
 
Random walks with absorbing or reflecting boundaries,     
or with memory, serve as important models in statistical   
physics, often admitting an exact analysis. Among the   
many examples are equilibrium models for polymer adsorption   
\cite{Barber,weiss,dba-havlin,Bell} and absorbing-state phase   
transitions \cite{Marro/Dickman}. Another motivation   
for the study of such problems is provided by the spreading   
of an epidemic in a medium with a long memory \cite{GCR}.  
  
In addition to the intrinsic interest of such an infection with         
memory, our study is motivated by the spread of activity in  
models exhibiting an infinite number of absorbing configurations 
(INAC), 
 typified by the pair contact process \cite{Jensen,Munoz/JSP}.   
Anomalies in critical spreading for INAC, such as continuously variable  
critical exponents, have been traced to a long memory in the dynamics  
of the order parameter, $\rho$, due to coupling to an auxiliary field  
that remains frozen in regions where $\rho = 0$  
\cite{Munoz/JSP,Munoz/PRL}.  
INAC appears to be particularly relevant to the transition to spatiotemporal  
chaos, as shown in a recent study of a coupled-map lattice with  
``laminar" and ``turbulent" states, which revealed continuously  
variable spreading exponents \cite{Bohr}.  

Grassberger, Chat\'e and Rousseau \cite{GCR} proposed that  
spreading in INAC could be understood by studying a model with a  
{\it unique} absorbing configuration, but in which the spreading rate   
of activity into previously inactive regions is different  
than for revisiting a region that has already been active.   
In light of the anomalies found in spreading in models with INAC  
or with a memory, we are interested in studying the effect of   
such a memory on the scaling behavior in a model whose asymptotic  
behavior can be determined exactly.  Of particular interest  
is the survival probability $S(t)$ (i.e., not to have fallen into  
the absorbing state up to time $t$).  

In the present work, we review previous results
on survival 
of 
 random walks with memory, and
analyze the asymptotic behavior 
of 
a random walk subject to  
a hard movable reflector.
On the basis of an exact  
solution for the probability generating function, we obtain the  
decay exponent $\delta$.    

The balance of this paper is organized as follows.
Section \ref{review}  reviews previous results on                                               
variable survival exponents for random walks with memory.
In Sec. \ref{model} we  
define the hard reflector model and obtain a formal 
solution for the generating function.
An asymptotic analysis of this function is presented 
in Sec. \ref{asymp}, leading to an expression for 
the decay exponent $\delta$ in terms of the reflection
probability $r$.
In Sec. \ref{sec/numerical} we present exact numerical results  
for finite times (from iteration of the probability transfer matrix)  
that complement and extend the asymptotic analysis.  
Sec. \ref{sec/discussion}  
contains a brief summary and discussion.  
  
\section{Variable survival exponents in random walks with memory}
\label{review}

Relatively simple random walk problems often serve as reduced examples
of much more complicated many-body phemomena.  So it is with phase
transitions between an active and an absorbing state.  In these systems
\cite{Marro/Dickman,Hinrichsen} the stationary state of a Markov
process exhibits (in the infinite-size limit) a phase transition from a
frozen, inactive regime to one with sustained activity, as a
control parameter is varied.  Broadly speaking, the control parameter 
represents the reproduction rate of activity (A $\to$ 2A) relative
to its extinction (A $\to$ 0).  The simplest examples are the
contact process \cite{harris} and directed percolation (DP).

Consider now, instead of the stationary state, the {\it spread} of activity
from a localized source, in an infinite system.  In the subcritical
regime (for which the only stationary state is the inactive, absorbing
one), the 
initial 
activity decays (typically, exponentially fast),
while in the supercritical regime there is a finite probability for it
to spread indefinitely.  Just at the critical point, one finds a 
scale-invariant evolution: the survival probability $S(t)$,
the integrated activity $n(t)$, and the mean-square distance, $R^2(t)$,
of the activity from the initial seed, all follow asymptotic
power laws.

The question of survival arises naturally in the context of a random 
walk in the presence of an absorbing boundary.  
The survival probability $S(t)$ is the probability never to have visited
an absorbing boundary until time $t$.  The simplest example
is a one-dimensional random walk $x_t$ (in discrete time)
on the non-negative integers, with the origin absorbing, and $x_0 \geq 1$.
Let the walker jump, at each time step, to the right ($x_{t+1} = x_t + 1$)
with probability $p$, and the left ($x_{t+1} = x_t - 1$) with probability
$ q = 1-p$.  If $p < q$ then the survival probability decays exponentially,
while for $p > q$ it approaches (again exponentially) a nonzero value,
so that the walker has a finite probability to escape to infinity.
In the absence of drift ($p = q = 1/2$), 
$S(t) \sim t^{-1/2}$ for
large $t$; associated with this power-law decay is an infinite mean lifetime.
In the analogy with an absorbing-state phase transition, $p=1/2$ evidently
marks the transition, with extinction certain for $p \leq 1/2$,
and a finite asymptotic survival probability for $p > 1/2$.  The
same qualitative situation holds in the contact process, starting for
example from a single active site \cite{Marro/Dickman}.

The analogy is in fact {\it exact} for the rather special case of
{\it compact directed percolation} (CDP), in which active regions
are delimited by independent random walks that annihilate on contact.
(CDP is a particular limit of the Domany-Kinzel model \cite{DK}.)
In this case `drift' corresponds to a tendency of the walkers
at the boundaries of an active region to approach, or separate 
from, one another; the critical point corresponds to zero drift, or unbiased 
random walkers ($p=1/2$).  Initializing CDP with a finite interval 
(say, 1, 2,...,m) of active sites, is equivalent to placing random 
walkers ($x_t$ and $y_t$) at 0 and $m+1$.  The active region at any subsequent time 
corresponds to the interval between the walkers $x_t$ and $y_t$; 
the process ends when the two meet.  To make the analogy between
CDP and a random walk with the origin absorbing complete,
we may fix $x_t = 0$ for all times, so that only the right frontier of
the active region is free to fluctuate, while the left frontier is pinned,
as it were, at a wall.

Given the connection with phase transitions, we shall think of the
power law for the survival probabilty as defining a critical exponent,
and write $S(t) \sim t^{-\delta}$.  The main interest of the
examples discussed in this paper is that the 
exponent $\delta$ can be shown to {\it vary continuously} as a function
of a parameter.  This in turn may help to understand scaling in
more complex examples, such as the pair contact process \cite{Jensen},
for which a variable survival exponent has been reported numerically,
but cannot be established rigorously.

Random walk models exhibiting variable survival exponents fall in two classes.
In one, the position of the absorbing boundary is a given 
(deterministic) 
function of time.  A random walk in the presence of such a boundary
defines a nonstationary stochastic process.  The second class 
(which is our principal interest here) involves {\it memory}, either in the form 
of a reflector that moves when it encounters the walker, or of a
history-dependent step length.

We begin with a brief review of the first class.  In a highly readable paper,
Krapivsky and Redner \cite{KR} considered what happens when a random walker
on the line is subject to two absorbing boundaries, $R(t)$ and $L(t)$,
which are prescribed functions of time
(initially the walker is at the origin).
The absorbers are initially near the origin,
and follow $R(t) = a + (At)^{\alpha}$, $L(t) = - R(t)$.  For 
$\alpha > 1/2$, the absorbers rapidly leave the region explored by the
walker, which therefore enjoys a finite probability of survival as
$t \to \infty$.  If the absorbers are stationary we of course expect
$S(t)$ to decay exponentially; but if their motion is characterized 
by $0 < \alpha < 1/2$, this changes to a {\it stretched exponential}
decay, $S \sim \exp[-t^{1-2\alpha} ]$.  
For $\alpha = 1/2$, the `safe'
region expands with the same power law as the region explored by the walker.
The result is power-law decay of the survival probability, with a
variable exponent 
which depends on $A$.  

The problem studied by Krapivsky and Redner may be pictured as a
random walk confined to a parabola in the $t - x$ plane, whose
equation (for $t \geq 0$) is $x = \pm (At)^{\alpha}$.  When
$\alpha = 1/2$, as noted, there is power-law decay of $S(t)$,
with a nonuniversal exponent.  Similar conclusions apply for DP
and for directed self-avoiding walks \cite{turban}, and for CDP
\cite{odor},
when these processes are confined to a fixed parabola.

We now turn to studies of a random walk
subject to some special condition when it enters virgin territory,
i.e., when it attempts to visit a site for the first time.  Here there is
no fixed confining boundary; the condition depends on the history
of the walk.  But since the region explored by the walker grows
$\propto t^{1/2}$, it effectively creates 
its 
own parabola.

It was recently shown \cite{rwmpr} that an
unbiased random walk on the nonnegative  
integers, with the origin absorbing,
exhibits a continuously variable exponent $\delta$  
when subject to a mobile, partial reflector.  The latter is initially  
one site to the right of the walker.  Each time the walker steps  
onto the site occupied by the reflector, it is reflected one step  
to the left with probability $r$ (it remains at its new  
location with probability $1\!-\!r$); in either case, the reflector   
is pushed forward one site in this encounter.  The survival exponent  
$\delta = (1\!+\!r)/2$ in this process    
\cite{rwmpr}. Since the reflector effectively 
records the {\it span} of the walk  
(i.e., the rightmost site yet visited), its interaction with the  
walker represents a memory.  
We shall refer to this model as the {\it soft reflector}, to highlight
the fact that the refector obligingly moves forward, even if the
walker is reflected back.  Note that in this case $\delta$ never exceeds
unity.

The analysis of the random walk with soft reflector was subsequently extended
to compact directed percolation \cite{cdppr}.  The active region initially consists of a
single site (the origin), and, as already noted, is bounded by a pair
of independent, unbiased random walkers, originally at $x=0$ and $x=1$.  
The two
walkers are subject to movable partial (``soft") reflectors,
such that the walker on the right is reflected toward the left
and vice-versa.  The results
for the survival exponent are qualitatively similar to those
for the single walker, but $\delta$ now varies between 1/2 and 1.160
as the reflection probability $r$ varies between zero and one.  The
results (coming in this case from iteration of the transition matrix,
rather than from an asymptotic analysis of the generating function),
are well fit by the simple expression $\delta = 1/2 + 2r/3$; small but
significant deviations from this simple formula are found, however.
CDP with reflectors has so far defied exact analysis, and the reason for the
specific value $\delta = 1.160$ for $r=1$ is not understood.

Most recently, the methods developed in Ref. \cite{rwmpr}
were applied to a one-dimensional random walk with  
memory of a different form: if the target site $x$ lies 
in the region that has been visited before
(that is, if $x$ itself has been visited, or lies between
two sites that have been visited), 
then the step length is $v$; otherwise the step length is $n$. 
In this case one finds $\delta = v/2n$ \cite{rwmem}.  Thus
$\delta$ can take any rational value between zero and infinity.

With this background we may now describe the problem to be analyzed
here as a random walk subject to a {\it hard} partial, movable reflector.
This is because the reflector now moves forward if and only if the
walker succeeds in occupying the new position; when the walker is reflected,
the reflector maintains its position.  This, as will be shown, can lead to much
larger values of $\delta$ than in the soft reflector case.

Before embarking on the technical discussion, we summarize our approach,
as developed in Refs. \cite{rwmpr} and \cite{rwmem}.  After formulating the
problem, we enlarge the state space so that the process becomes
Markovian in the expanded representation \cite{vanKampen}.  We then write down the
equation of motion for the probability distribution and its associated boundary
and initial conditions.  Since these are discrete models, the equation of
motion corresponds to a set of {\it difference equations}, first-order
in time, and second-order in space.  It is convenient to eliminate the
time variable by passing to a generating function $\hat{P}(z)$
(effectively, a discrete Laplace transform).  Using separation of variables,
we obtain a formal solution for the generating function.  Finally,
the asymptotic long-time behavior is found by studying the
generating function for the survival probability in the limit $z \to 1$.

\section{Model}  
\label{model}

Consider an unbiased, discrete-time random walk on the nonnegative   
integers, with the origin absorbing. We denote the position of   
the walker at time $t$ by $x_{t}$, with $x_{0}=1$.  
The movement of the walker is affected by the presence of a
movable partial reflector, whose position is denoted by $R_t$,
with $R_0 = 2$.  At each time step the walker hops from its
current position $x_t$ to either $x_t +1$ or $x_t -1$ with
probabilities of 1/2.  If, however, $x_t +1 = R_t$, the walker
is reflected back to $x_t$ with probability $r$, and remains
at $x_t +1$ with probability $\overline{r} \equiv 1\!-\!r$; in the latter case the reflector
simultaneously moves to $R_t +1$.  Summarizing, the transition probabilities
for the walker are

\begin{equation}  
\label{trp1}
x_{t+1}   
=  
\left \{  
\begin{array}{ll}  
x_{t} - 1 & \mbox{, w.p. } 1/2 \\  
x_{t} + 1 & \mbox{, w.p. } 1/2 
\end{array}  
\right.  
\end{equation}   
in case $x_{t} \leq R_t -2$.  When $x_t = R_t -1$, we have instead

\begin{equation}  
\label{trp2}
x_{t+1}   
=  
\left \{  
\begin{array}{ll}  
x_{t} - 1 & \mbox{, w.p. } 1/2 \\  
x_{t} + 1 & \mbox{, w.p. } \overline{r}/2 \\
x_t       & \mbox{, w.p. } r/2
\end{array}  
\right.  
\end{equation}   
The position of the reflector at any moment is given by
$R_t = 1 + \max_{t' \leq t} \{x_{t'} \}$.

Although the process $x_{t}$ is non-Markovian (since the   
transition probability into a given site depends on whether    
it has been visited previously),
we can define a Markov process by expanding the state space
to include   
the variable $y_t \equiv R_t - 1 = \max_{t' \leq t} \{x_{t'} \}$.  
The state space
$E \subset {\Bbb Z}^{2}$   
is given by by   
\[   
E  
=  
\{  
(x,y) \in {\Bbb Z}^{2}:  
x \geq 0, \;\;  
y \geq 1, \;\;  
x \leq y   
\} , 
\]  
as represented in Fig. 1.  

The probability distribution
$P(x,y,t)$ follows the evolution equation  
\begin{equation}  
\label{evolution}  
P(x,y,t+1) = \frac{1}{2}P(x+1,y,t) + \frac{1}{2}P(x-1,y,t),  
\hspace{1cm} \mbox{for $x<y-1$},  
\end{equation}  
with $P(x,y,0)=\delta_{x,1}\delta_{y,1}$. Eq. (\ref{evolution}) is 
subject to two boundary conditions. The first is 
the absorbing condition for $x \leq 0$
\begin{equation}  
\label{bc at x=0}  
P(x,y,t) = 0, \hspace{1cm} \mbox{for $x \leq 0$.}  
\end{equation}  
The second applies along the diagonal $x\!=\!y$.    
Defining
$D(y,t) \equiv P(y,y,t)$, we have 
\begin{equation}  
\label{bc at x=y}  
D(y,t+1) = \frac{1}{2}P(y \!-\!1,y,t) + \frac{\overline{r}}{2}\;D(y \!-\!1,t) + 
\frac{r}{2} D(y,t),  
\hspace{1cm} \mbox{for $y \geq 2$}.  
\end{equation}  
For $y=1$ the equation is simply $D(1,t+1) = (r/2) D(1,t)$, and since
$D(1,0) = 1$, one has
$D(1,t) = (r/2)^t$.  Finally, for $x=y-1$,
\begin{equation}  
\label{evolution1}  
P(y \!-\!1,y,t+1) = \frac{1}{2}P(y \!-\!2,y,t) + \frac{1}{2}D(y,t),  
\hspace{1cm} \mbox{for $x<y-1$},  
\end{equation}

We next introduce the generating function: 
\begin{equation}  
\label{generating}  
\hat{P}(x,y,z) = \sum_{t=0}^{\infty} P(x,y,t) \; z^{t}.  
\end{equation}  
Multiplying Eqs. (\ref{evolution}), (\ref{evolution1}) 
by $z^{t}$, summing over $t$ and shifting the sum index  
where necessary, one finds that the generating   
function satisfies   
\begin{eqnarray}  
\label{evgen}  
\frac{1}{z}\hat{P}(x,y)   
&=&   
\frac{1}{2}\hat{P}(x+1,y)   
+   
\frac{1}{2}\hat{P}(x-1,y),  
\hspace{1cm} \mbox{for $x \leq y-2$}\\  
\label{evgen1}  
\frac{1}{z}\hat{P}(y \!-\!1,y)   
&=&   
\frac{1}{2}\hat{D}(y)   
+   
\frac{1}{2}\hat{P}(y \!-\!2,y),  
\hspace{2.1cm} \mbox{for $x=y-1$},  
\end{eqnarray}  
(we drop the argument $z$ for brevity), where    
$\hat{D}(y)$ is defined by an expression analogous   
to Eq. (\ref{generating}). The   
initial condition implies $\hat{D}(1)=(1-zr/2)^{-1}$; the   
boundary conditions are  
\begin{equation}  
\label{bcgen}  
\hat{P}(0,y)  = 0, 
\end{equation}
and
\begin{equation} 
\label{bcgen1}  
\frac{1}{z}\hat{D}(y)   
 =    
\frac{1}{2}\hat{P}(y \!-\!1,y) + \frac{\overline{r}}{2}\hat{D}(y \!-\!1) 
+\frac{r}{2} \hat{D}(y),  
\hspace{1.5cm} \mbox{for $y \geq 2$}.  
\end{equation}
  
Eq. (\ref{evgen}) relates $\hat{P}$ at different values of $x$,
for the same $y$.  Specifically, on the interior of
each line of constant $y$, $\hat{P}$ satisfies a diffusion equation,
with a source at $x=y$ and a sink at $x=0$.  It is therefore natural
to attempt separation of variables,
\begin{equation}
\hat{P}(x,y) = \hat{A}(x)\hat{B}(y) \;.  
\label{sepvar}
\end{equation}
Inserting this expression in
Eq. (\ref{evgen}) one obtains
\begin{equation}
\frac{1}{z}\hat{A}(x) - \frac{1}{2}\hat{A}(x-1) - \frac{1}{2}\hat{A}(x+1)
=0      \;,
\label{diffx}
\end{equation}
with $A(0)=0$.  The solution satisfying this boundary condition is
\begin{equation}
\hat{A}(x) = \lambda^x - \lambda^{-x} \;, 
\label{solA}
\end{equation}
with 
\begin{equation}
\lambda
=
\frac{1}{z}
+
\sqrt{\frac{1}{z^{2}} -1} \;.
\label{lambda}
\end{equation}

Our next task is to determine $\hat{B}(y)$; for this we require
a relation between generating functions with different arguments
$y$.  Relations of this kind arise along the diagonal, but involve
the function $\hat{D}(y)$, which we proceed to eliminate.
Combining Eqs. (\ref{evgen1}) and (\ref{bcgen1}), we find
\begin{equation}
Q(z) \hat{D}(y) = \frac{z\overline{r}}{2} \hat{D}(y \!-\!1) 
+ \frac{z^2}{4} \hat{P}(y \!-\!2,y)
\label{dd}
\end{equation}
where 
\begin{equation}
Q(z) = 1 -\frac{zr}{2} - \frac{z^2}{4} \;.
\label{defQ}
\end{equation}
Equation (\ref{bcgen1}) may also be written as
\begin{equation}
\left(1 - \frac{zr}{2} \right) \hat{D}(y) 
- \frac{z \overline{r}}{2} \hat{D}(y \!-\!1) 
= \frac{z}{2} \hat{P}(y \!-\!1,y) \;,
\label{dd1}
\end{equation}
We now multiply Eq. (\ref{dd}) for $y-1$ by $\beta \equiv z\overline{r}/(2-zr)$
and subtract the result from the same equation for $y$.  This
yields
\begin{equation}
Q \left[\hat{D}(y) - \beta \hat{D}(y\!-\!1) \right]
= \frac{z\overline{r}}{2} 
\left[\hat{D}(y\!-\!1) - \beta \hat{D}(y\!-\!2) \right]
 + \frac{z^2}{4} 
 \left[\hat{P}(y\!-\!2,y) - \beta \hat{P}(y\!-\!3,y\!-\!1) \right] \;.
\label{dd2}
\end{equation}
>From Eq. (\ref{dd1}) we have
\begin{equation}
\hat{D}(y) - \beta \hat{D}(y \!-\!1) = \frac{z}{2-zr} \hat{P}(y \!-\!1,y) \;,
\label{dd3}
\end{equation}
allowing us to eliminate $\hat{D}$ from Eq. (\ref{dd2}):
\begin{equation}
\frac{4Q}{z(2-zr)} \hat{P}(y \!-\!1,y)  - \hat{P}(y \!-\!2,y) =
\beta\left[ \frac{2}{z} \hat{P}(y \!-\!2,y-1)  -\hat{P}(y \!-\!3,y-1) \right]  \;.
\label{Pdiag}
\end{equation}
Inserting Eq. (\ref{sepvar}) one readily finds a recursion relation
for $\hat{B}$:
\begin{equation}
\frac{\hat{B}(y)}{\hat{B}(y \!-\!1)} =
\frac{z\overline{r}[2\hat{A}(y \!-\!2) - z\hat{A}(y \!-\!3)]}
{(4-2zr-z^2)\hat{A}(y \!-\!1) - z (2-zr)\hat{A}(y \!-\!2)}    \;.
\label{recrel}
\end{equation}
Given $\hat{B}(1) = \hat{D}(1)/\hat{A}(1)$ with $\hat{D}(1)$
and $\hat{A}(y)$ as found above, Eqs. (\ref{sepvar}), (\ref{solA}),
and (\ref{recrel}) represent a complete formal solution for the
generating function $\hat{P}(x,y)$.

\section{Asymptotic analysis}
\label{asymp}

Our goal is to find the survival probability $S(t)$ for large $t$.
This can be found analysing the associated generating function,
\begin{equation}
\hat{S}(z) = \sum_{t=0}^\infty S(t)z^t
\label{defshat}
\end{equation}
in the limit $z \to 1$.  Specifically, if $S(t) \sim t^{-\delta}$,
then the radius of convergence of $\hat{S}(z)$ is $|z|=1$, and the singular 
behavior of the generating function as $z \to 1$ determines $\delta$.
Indeed, in this case, with $z = 1- \epsilon$, we have
\begin{eqnarray}
\nonumber
\hat{S} & \simeq & \sum_{t=1}^\infty t^{-\delta} (1-\epsilon)^t
\\
\nonumber
 & \simeq & \int_1^\infty dt \; t^{-\delta} \exp [-t |\ln (1-\epsilon)|] 
\\
 & \simeq & \epsilon^{\delta - 1} \Gamma (1-\delta) \;,
\label{sasympt}
\end{eqnarray}
so that the scaling
exponent $\delta$ can be read off from the power-law dependence
of $\hat{S}$ on $\epsilon = 1 - z$ as $\epsilon \to 0$.  This simplifies
considerably the determination of the long-time asymptotic 
behavior of $S(t)$.

The generating function $\hat{S}$ has two contributions, coming from
the ``interior" ($x < y$) and the diagonal:
\begin{eqnarray}
\nonumber
\hat{S} &  = & \sum_{y=1}^\infty \sum_{x=1}^{y-1} \hat{P}(x,y)
+ \sum_{y=1}^\infty  \hat{D}(y)
\\
 & \equiv & \hat{S}_P + \hat{S}_D \;.
\label{SpSd}
\end{eqnarray}
Using Eq. (\ref{dd3}) one readily shows that
\begin{equation}
\hat{S}_D = \frac{z}{2-z} \sum_{y=1}^\infty \hat{P}(y \!-\!1,y) < \hat{S}_P \;,
\label{ineq}
\end{equation}
so that it suffices to analyze the behavior of $\hat{S}_P$.

Consider
\begin{equation}
\hat{S}_P = \sum_{y=1}^\infty \hat{B}(y) \sum_{x=1}^{y-1} \hat{A}(x) .
\label{Spexpl}
\end{equation}
The sum over $x$ can be evaluated as
\begin{eqnarray}
\nonumber
\sum_{x=1}^{y-1} \hat{A}(x)
&=& \frac{\lambda^y -1}{\lambda -1} - 
\frac{\lambda^{-y} -1}{\lambda^{-1} -1} 
\\
& \simeq & 
\frac{4}{\Lambda} \sinh^2 \frac{\Lambda y}{2}
\label{sumA}
\end{eqnarray}
where in the last step we used 
$\Lambda \equiv \ln \lambda \simeq  \sqrt{2\epsilon}$ 
as $\epsilon \to 0$.
 
To evaluate $\hat{S}_P$ we also require an expression for 
$\hat{B}(y)$ in the limit $\epsilon \to 0$;
this can be obtained from Eq. (\ref{recrel}).  We begin 
by setting all explicit factors of $z$ equal to unity,
since the ${\cal O}(\epsilon)$ corrections thereby discarded
do not contribute to the singular behavior of $\hat{S}_P$.
The singular contributions in fact originate from the
functions $\hat{A}$, through their dependence on $\lambda$.
Writing $\hat{A}(y) = 2 \sinh \Lambda y$, we therefore have 
\begin{equation}
\frac{\hat{B}(y)}{\hat{B}(y \!-\!1)} \simeq
\frac{\overline{r}[2\sinh \Lambda (y \!-\!2) - \sinh \Lambda (y \!-\!3)]}
{(3-2r) \sinh \Lambda(y \!-\!1) - (2-r) \sinh \Lambda (y \!-\!2)}    \;.
\label{rec1}
\end{equation}
Using the identity $\sinh (a+b) = \sinh a \cosh b + \sinh b \cosh a$,
and neglecting terms ${\cal O}(\Lambda^2)$, we obtain 
\begin{equation}
\frac{\hat{B}(y)}{\hat{B}(y \!-\!1)} \simeq
\frac{\tanh \Lambda y - \Lambda}
{\tanh \Lambda y + \Lambda/\overline{r}}    \;.
\label{rec2}
\end{equation}

For $y \leq y_0 = [2/\overline{r}] +1 $ (here $[...]$ denotes the
integer part of its argument), we can write, for small $\Lambda$
\begin{equation}
\hat{B}(y) = \hat{B}(1) \prod_{k=2}^y \frac{k-1}{k+1/\overline{r}}
= C \hat{B}(1)        \;,
\label{yleqy0}
\end{equation}
where $C$ depends on $r$ and $y$ but is independent of $\Lambda$.
We shall in fact discard the contribution due to $y < y_0$ in
$\hat{S}_P$.  The reason is that the contribution to the survival
probability from any fixed, finite set of transient states must
decay exponentially at long times, and so will not affect our
result for the scaling exponent.  

Noting that $\hat{B}(1) = \hat{D}(1)/\hat{A}(1) \propto 1/\Lambda$,
we have, for $y > y_0$,
\begin{equation}
\hat{B}(y) = \frac{C}{\Lambda}  \prod_{k=y_0}^y 
\frac{\hat{B}(k)}{\hat{B}(k\!-\!1)}
       \;,
\label{ygty0}
\end{equation}
where $C$ is a constant.  Since all terms have $k \geq 2/\overline{r}$,
we may use Eq. (\ref{rec2}) to write, with
$\phi_k \equiv \tanh \Lambda k$,
\begin{eqnarray}
\nonumber
\ln \frac{\hat{B}(y)}{\hat{B}(y_0)} & \simeq & \sum_{k=y_0}^y 
\ln \frac{1-\Lambda/\phi_k}{1+\Lambda/(\overline{r}\phi_k)}
\\
& \simeq & -\left(1 + \frac{1}{\overline{r}} \right) \Lambda
\sum_{k=y_0}^y \frac{1}{\phi_k}
       \;.
\end{eqnarray}
Approximating the sum by an integral we find
\begin{equation}
\ln \frac{\hat{B}(y)}{\hat{B}(y_0)}  \simeq  
 \left(1 + \frac{1}{\overline{r}} \right) 
\ln \frac{\sinh \Lambda y_0}{\sinh \Lambda y}
       \;.
\label{Bofy}
\end{equation}

Now, inserting Eqs. (\ref{sumA}) and (\ref{Bofy}) in Eq. (\ref{Spexpl}),
the generating function for $\epsilon \to 0 $ is: 
\begin{equation}
\hat{S}_P   \sim  
 \Lambda^{1/\overline{r}-1} \sum_{y=y_0}^\infty
 \frac{\sinh^2 \Lambda y/2}{\sinh^{1+1/\overline{r}} \Lambda y}
       \;,
\label{SPasymp}
\end{equation}
where `$\sim$' denotes asymptotic proportionality as $\epsilon \to 0$.
Approximating, as before, the sum by an integral, we have
\begin{equation}
\hat{S}_P   \sim  
 \Lambda^{1/\overline{r}-2} \int_{\Lambda y_0}^\infty du
 \frac{\sinh^2 u/2}{\sinh^{1+1/\overline{r}} u}
       \;.
\label{SPint}
\end{equation}

Since $1+1/\overline{r} \geq 2$, the integral converges at its upper limit.
For $r < 1/2$, 
$1+1/\overline{r} <3 $ and the integral remains finite
as $\Lambda \to 0$.  Then
\begin{equation}
\hat{S}_P   \sim  
 \Lambda^{1/\overline{r}-2} \sim \epsilon^{1/(2\overline{r}) - 1} \;,
\label{caseI}
\end{equation}
so that the survival probability decays as $S(t) \sim t^{-\delta}$
with $\delta= 1/2\overline{r}$.  For $r=1/2$, the prefactor
in Eq. (\ref{SPint}) is independent of $\Lambda$ and
\begin{equation}
\hat{S}_P   \sim \int_{\Lambda y_0}^\infty \frac{du}{u} 
\sim - \ln (1-z)   \;.
\label{caseII}
\end{equation}
Expanding the logarithm, we find $\hat{S}(z) \sim \sum_n z^n/n$,
yielding directly $S(t) \sim t^{-1}$.
Finally, when $r > 1/2$, $1+1/\overline{r} > 3 $ and the integral
in Eq. (\ref{SPint}) contains two principal contributions: one
finite (due to the interval from say, 1, to infinity), the other
arising from the lower limit, and diverging as 
$\Lambda^{2-1/\overline{r}}$.  Combined with the prefactor
$\propto \Lambda^{1/\overline{r}-2} $ however, the latter contribution
is nonsingular, while the former is again proportional
to $\Lambda^{1/\overline{r}-2}$.

Summarizing, the asymptotic survival probability decays as a power
law,
\begin{equation}
S(t) \sim t^{-1/2\overline{r}}   \;,
\label{surv}
\end{equation}
which is the result we set out to prove.

\section{Numerical Results}
\label{sec/numerical}

The foregoing analysis provides the $t \to \infty$ scaling behavior
of the survival probability,
but does not indicate the rate of convergence to the asymptotic power
law.  To determine how the corrections to scaling decay, we iterate
the discrete time evolution equation for $P(x,y,t)$ directly.
In Fig. 2 we show the decay of $S(t)$ for reflection probability
$r=0.85$, corresponding to $\delta = 10/3$.  For very late times,
the graph indeed approaches a power law with the expected
exponent.   The approach is, however, extremely slow. 

Since our asymptotic analysis only retains the leading dominant term
in the long-time behavior of $S(t)$, we have no specific information
on correction to scaling terms.  It is easy to see, nonetheless, that corrections
$\propto t^{-1/2}$ will be generated, since $\Lambda \simeq  \sqrt{2\epsilon}
+ {\cal O}(\epsilon)$.  In fact, we are able to fit the long-time evolution
of the survival probability by adding a suitable ${\cal O}(t^{-1/2}) $ term
to the power law, but further terms ($\sim t^{-1})$, etc.) are required for 
intermediate times.  

We have found a particularly simple transformation of variable that appears to
take the dominant correction to scaling into account.  It consists in defining a
shifted time variable
\begin{equation}
T = t + bt^{1/2} ,
\label{deftprime}
\end{equation}
with parameter $b$ adjusted to make the graph of $S$ versus $T$ (on log scales)
as linear as possible.  Fig. 3 shows the survival probability data for $r = 1/2$,
2/3, 0.75 and 0.85 versus $T$ (the corresponding $b$ values are 1.754, 4.167,
7.042, and 16.67).  In each case the 
numerical 
data (points) follow the
modified power law,
\begin{equation}
S(T) = A T^{-\delta} ,
\label{mpl}
\end{equation}
to very high precision.  [Here $A$ is an amplitude determined by extrapolating
$T^{\delta}S(T)$ to $T \to \infty$.]  The numerical data appear to converge
rapidly (faster than a power law) to the fit.  While this `shifted time' analysis 
is for the moment without theoretical basis, it clearly confirms the asymptotic
power laws found analytically, and suggests a simple form for describing slowly
decaying corrections to scaling.

The slowly decaying correction to scaling would likely frustrate
efforts to extract 
the correct long-time behavior from
simulations.  Looking at Fig. 2, we see that the asymptotic power
law is barely evident when $S(t)$ has decayed to $e^{-15}$. To 
obtain even marginally useful simulation data in this situation 
we would need to perform $\geq 10 e^{15} \simeq 3 \times 10^7$ 
independent realizations of the process, extending to a maximum time
of about 2000 steps.  This is feasible for a simple random walk, but
becomes a computational challenge for a many-particle system.
Thus, if lattice models such as the PCP behave in a manner analogous
to what is found for the random walk with a hard reflector, it
will be very difficult to confirm power-law scaling in simulations.
Data for limited times (or limited samples) may well give the
impression of faster than exponential decay of $S(t)$.

\section{Discussion}
\label{sec/discussion}

We have reviewed examples of confined random walks, and random walks with memory,
that lead to a continuously variable scaling exponent for the survival probability,
and analysed in detail the `hard reflector' case.  The latter problem appears to
be particularly relevant to spreading in the pair contact process, since modification
of the background density of isolated particles can only occur when activity
invades a previously inactive region \cite{Jensen}.  The strong correction
to scaling found numerically for the hard reflector is reminiscent of the slow convergence
(interpreted as faster than power-law decay in Ref. \cite{GCR}), found in spreading
studies of the PCP.

Several interesting issues remain open.  First, the nature of correction to scaling terms
needs to be investigated using a more complete asymptotic expansion of the
generating function.  Second, one would like to understand the exponent values for CDP
(obtained numerically in Ref. \cite{cdppr}) on the basis of the generating 
function approach.  Finally, extension of any of the models discussed here to two or more
dimensions promises to be a difficult but potentially fascinating challenge.

\acknowledgments
We thank Deepak Dhar and Miguel A. Mu\~noz for helpful discussions. 
R.D. and F.F.A. acknowledge financial support from CNPq (Brazil); 
D. b-A. acknowledges support of
NSF (USA) under grant PHY-0140094.

\newpage 
\noindent{Figure Captions}  
\vspace{1em} 

\noindent Fig. 1. Random walk subject to hard reflector: 
transitions in the $x$-$y$ plane. 
\vspace{1em} 
  
\noindent Fig. 2. 
Decay of survival probability $S(t)$ for reflection 
probability $r=0.85$ (solid curve); 
the slope of the straight line is -10/3.
\vspace{1em} 

\noindent Fig. 3. 
Survival probability $S$ as a function of the shifted
time variable $T$, for reflection 
probabilities $r=1/2$, 2/3, 3/4 and 0.85 (data points); 
the straight lines have slopes of -1, -3/2, -2, and -10/3.


\begin{thebibliography}{99}  

\bibitem{Barber}  
M. N. Barber and B. W. Ninham,   
{\it Random and Restricted Walks},   
Gordon and Breach, New York, 1970.  
  
\bibitem{weiss}
     G. H. Weiss,
     {\it Aspects and Applications of the Random Walk},
     (North Holland, Amsterdam, 1994).

\bibitem{dba-havlin}
     D. ben-Avraham and S. Havlin,
     {\it Diffusion and Reactions in Fractals and Disordered Systems},
     (Cambridge University Press, Cambridge, 2000).


\bibitem{Bell}
K. De'Bell and T. Lookman, 
Rev. Mod. Phys. {\bf 65}, 87 (1993).

  
\bibitem{Marro/Dickman}  
J. Marro and R. Dickman,   
{\it Nonequilibrium Phase Transitions in Lattice Models},   
Cambridge University Press, Cambridge, 1999.  
  
  
\bibitem{GCR}  
P. Grassberger, H. Chat\'{e}, G. Rousseau,   
Phys. Rev. E {\bf 55}, 2488 (1997).  
  
  
\bibitem{Jensen}  
I. Jensen, Phys. Rev. Lett. {\bf 70}, 1465 (1993);  
I. Jensen and R. Dickman, Phys. Rev. E {\bf 48}, 1710 (1993).  
 
 
\bibitem{Munoz/JSP} 
     M. A. Mu\~noz, G. Grinstein, and R. Dickman,  
     J. Stat. Phys. {\bf 91}, 541 (1998). 
      

\bibitem{Munoz/PRL} 
     M. A. Mu\~noz, G. Grinstein, R. Dickman, and R. Livi,  
     Phys. Rev. Lett. {\bf 76}, 451 (1996). 
      
     
\bibitem{Bohr}  
     T. Bohr, M. van Hecke, R. Mikkelsen, and M. Ipsen, 
     Phys. Rev. Lett. {\bf 86}, 5482 (2001). 
  
\bibitem{Hinrichsen}
	    H. Hinrichsen, 
	    Adv. Phys.  49, 815 (2000).

\bibitem{harris} 
	T. E. Harris, 
	Ann. Prob.  2 (1974) 969.

\bibitem{DK}
	E. Domany and W. Kinzel,
	Phys. Rev. Lett. {\bf 53}, 311 (1984).
  
\bibitem{KR}
	P. Krapivsky and S. Redner,
	Am. J. Phys. {\bf 64}, 546 (1996).
 

\bibitem{turban}
     L. Turban,
     J. Phys. A {\bf 25}, L127 (1992);
     C. Kaiser and L. Turban,
     {\it ibid.}, {\bf 27}, L579, (1994).

\bibitem{odor} 
     G. \'Odor and N. Menyh\'ard,
     Phys. Rev. E {\bf 61}, 6404 (2000).

\bibitem{rwmpr}  
	   R. Dickman and D. ben-Avraham,   
	   Phys. Rev. E {\bf 64}, 020102 (2001).  

\bibitem{cdppr}  
	   R. Dickman and D. ben-Avraham,   
	   J. Phys. A {\bf 35}, 7983 (2002).  

\bibitem{rwmem}  
	   R. Dickman, F. F. Araujo Jr., and D. ben-Avraham,   
	   Phys. Rev. E {\bf 66}, 051102 (2002).  

\bibitem{vanKampen}  
N. G. van Kampen,   
{\it Stochastic Processes in Physics and Chemistry},   
North-Holland, Amsterdam, 1992.  
 
  
\bibitem{Montroll/1965}  
E. W. Montroll, and G. H. Weiss,   
J. Math. Phys. {\bf 6}, 167-181 (1965).  

  
\bibitem{Spitzer}  
F. Spitzer,   
{\it Principles of Random Walk},   
Springer, New York, 1976.  
  
 
  
\end{thebibliography}
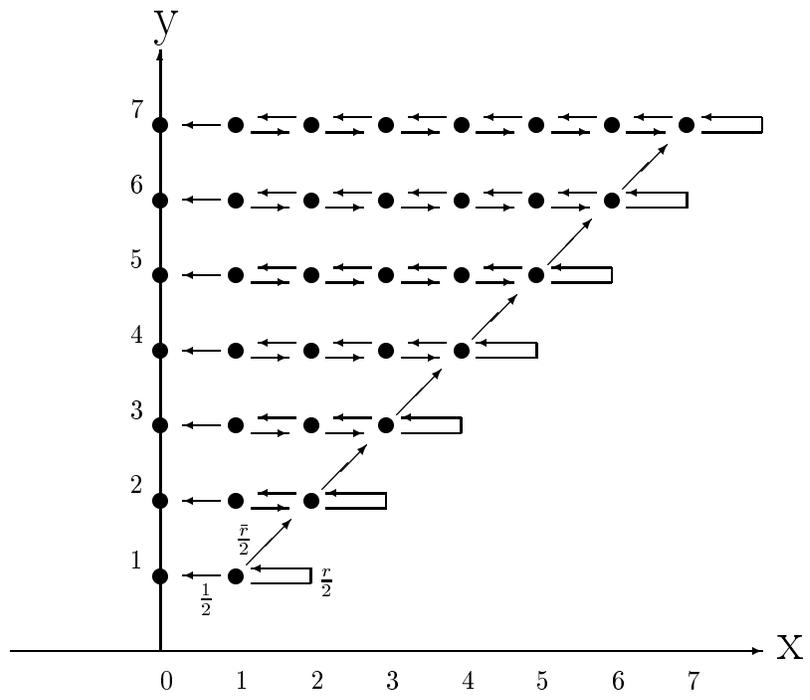
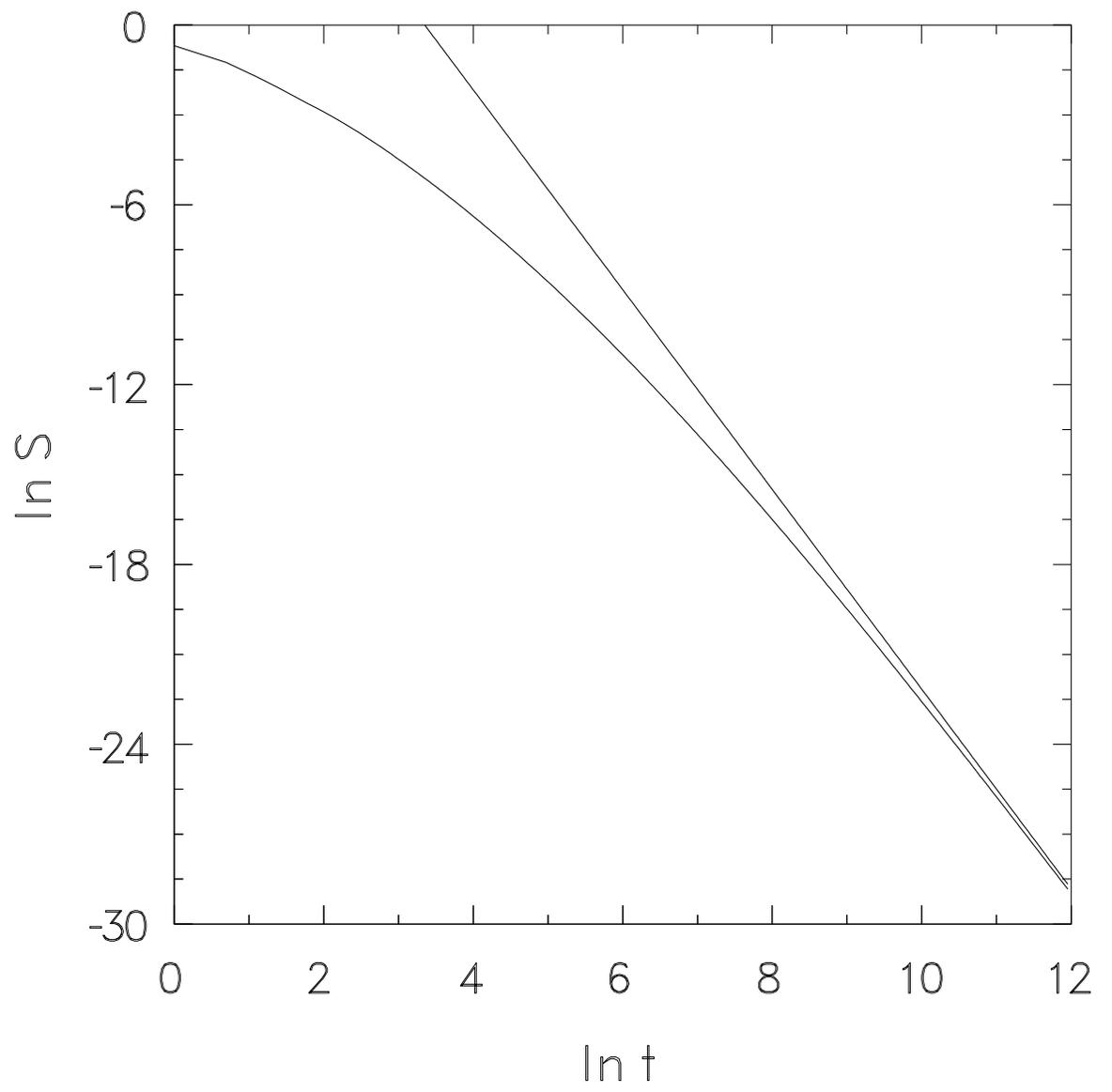
\end{document}